\documentclass[10pt]{article}
\usepackage{graphicx}

\newcommand{\km}{\,\mbox{km}\,\mbox{s}^{-1}}

\begin{document}
\textheight=23cm
\textwidth=16.5cm
\topmargin=-1.5cm
\hoffset=-1.5cm

\begin{center}
{\large \bf
Long-slit and Fabry-Perot spectroscopy of \\
collisional ring galaxy Arp~10}
\end{center}

\vskip 0.5cm

{\large Moiseev A.V.$^1$, Bizyaev D.V.$^{2,3}$, Vorobyov, E. I.$^{4,5}$ }
\medskip

{\footnotesize
(1) Special Astrophysical Observatory, Russian Academy of Sciences

(2) University of Texas at El Paso

(3) Sternberg Astronomical Institute, Moscow, Russia

(4) University of Western Ontario, London ON, Canada

(5) Institute of Physics, Rostov-on-Don, Russia
}

\vskip 0.5cm

\centerline{\rm ABSTRACT}
\medskip

We present results of Fabry-Perot and long-slit spectroscopy of the peculiar
galaxy Arp~10. The ionized gas velocity field shows evidence for significant
radial motions in both outer and inner galactic rings. Long-slit spectroscopy
reveals gradients of age and metallicity of stellar population in agreement
with the propagating nature of star formation in the galaxy. We present strong
evidence that a small ``knot'' at 5 arcsec from the center of Arp~10 is its
dwarf elliptic satellite, the most probable ``intruder'' responsible for
triggering the expanding rings in Arp~10.

\section{Introduction}

Arp~10 appeared in the Arp Atlas of Peculiar Galaxies [1] as an RN galaxy,
i.e. a galaxy containing a ring sructure and a bright nucleus. Further high
resolution studies of Arp~10 ([2],[3]) have shown that it actually has two
rings (the inner and outer one), nucleus, and extended outer arc. There exist
a contraversy about the origin of such a complicated structure. According to
[3], the system of rings and arcs in Arp~10 is formed by a direct face-on
collision with one of Arp~10 companions, in which case Arp~10 represents a
classical collisional ring galaxy. In the later paper [4] the same authors
argue that Arp~10 is formed as a result of merging of a companion with the
disk of Arp~10, in which case Arp~10 is a merging system. The nearby companion
was rejected as an intruder candidate and a bright unresolved knot at $5''$
from the center of the galaxy was proposed as a possible candidate to the real
intruder ([4]). The oxygen abundance of one half of the solar was estimated
for a few bright knots in the outer ring ([5]).

\section{Observations}

We conducted the complex spectral study of Arp~10 with the help of scanning
Interferometer Fabry-Perot (IFP) as well as a long-slit spectroscopy.
Observations were carried out with the multimode focal reducer SCORPIO ([6])
at 6-m telescope (Special Astrophysical Observatory, Russian Academy of
Sciences) in September and November 2003. The main goals of this study are to
trace and to model the propagation of induced wave of star formation through
the disk of Arp~10, to estimate the age of the rings and to identify the
intruder.

\begin{figure}
\includegraphics[width=13cm,angle=0]{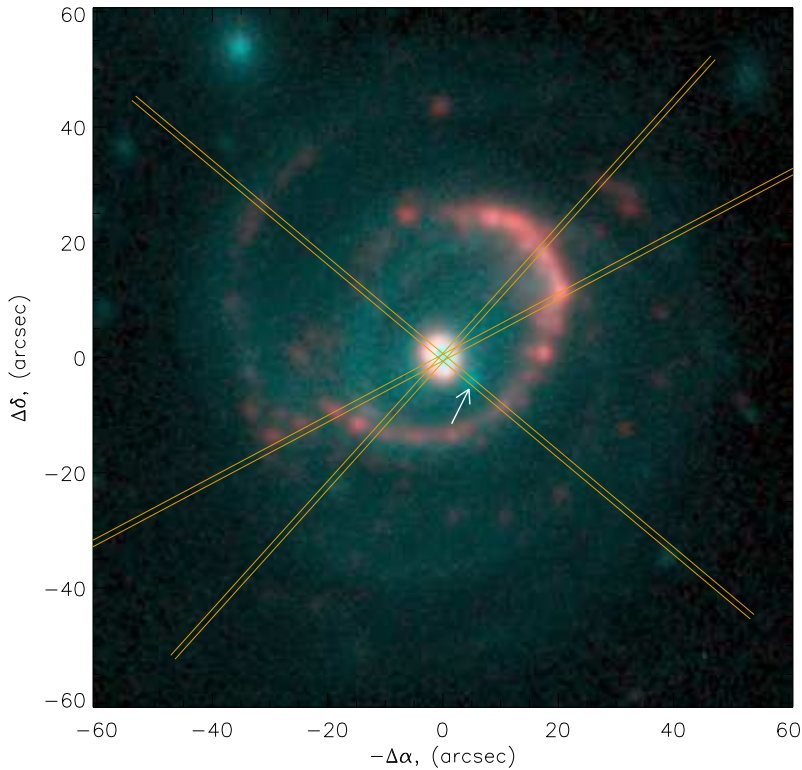}
\caption{ $H_\alpha$ emission map of Arp~10 (red) superimposed on the B-band
image (blue). The yellow lines indicate positions of the slits. The white
arrow marks ``the intruder''. }
\end{figure}

The $H_\alpha$ emission map is shown in Fig. 1. Two star forming rings
and the outer arc are clearly seen, the latter looks like debris of a spiral arm. It
implies that pre-collisional Arp~10 was a large spiral galaxy and it had a
non-negligible stellar population which formed spiral arms.

\begin{figure}
\includegraphics[width=9cm,angle=0]{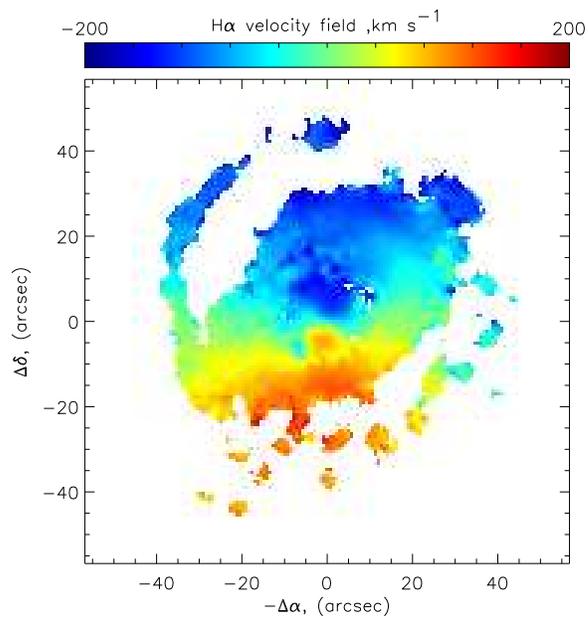}
\caption{ The velocity field of Arp~10 in $H_\alpha$ }
\end{figure}

\begin{figure}
\includegraphics[width=9cm,angle=0]{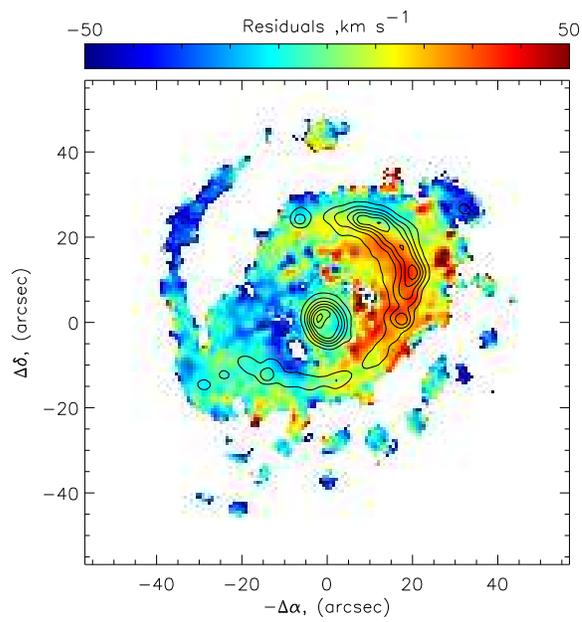}
\caption{ Residuals after subtracion of circular motion from Fig.2 }
\end{figure}

The  $H_\alpha$ velocity field of Arp~10 was obtained via 7320 seconds
integration with IFP in 32 spectral channels. It is shown in Fig. 2. We fit a
model velocity field taking into account the regular rotation and projection
effects. The radial velocities in the central part of Arp~10 are shifted
systematically against all other regions in the galaxy. This
feature is most probably caused by systematic vertical motions of the ionized gas around
the inner ring. Figure~3 shows the velocity residuals map which reveals the presence of
an expanding outer ring. The superimposed isophotes in Fig.~3 show the $H_\alpha$ emission
which seems to coinside with the highest values of residuals. The expansion velocity
of the NW part of the outer ring exceeds 100 $\km$ (in the plane), whereas it attain only 30
$\km$ at the SE part. Therefore, the asymmetric shape of the outer ring may
be caused  by a systematic difference in the ring expansion velocity.

\begin{figure}
\includegraphics[width=12cm,angle=0]{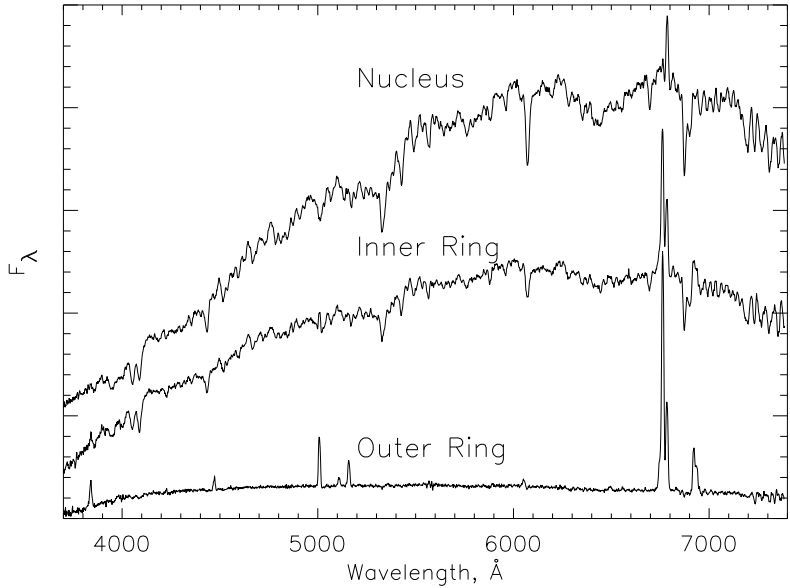}
\caption{ Spectra of nucleus, inner and outer rings of Arp~10.}
\end{figure}

Three long-slit cuts were made through the nucleus of the galaxy and through
the knot (see Fig. 1 for the cut positions). The spectral resolutions were 5 and
11 \AA/pixel, and the overall total integration time was 13500 sec. The spectra
of nucleus, inner and outer rings are shown in Fig.~4. They reveal a bulk of
young stellar population in the rings in contrast to the center of Arp~10, where the old
population dominates.

\begin{figure}
\centerline{
\includegraphics[width=5cm,angle=0]{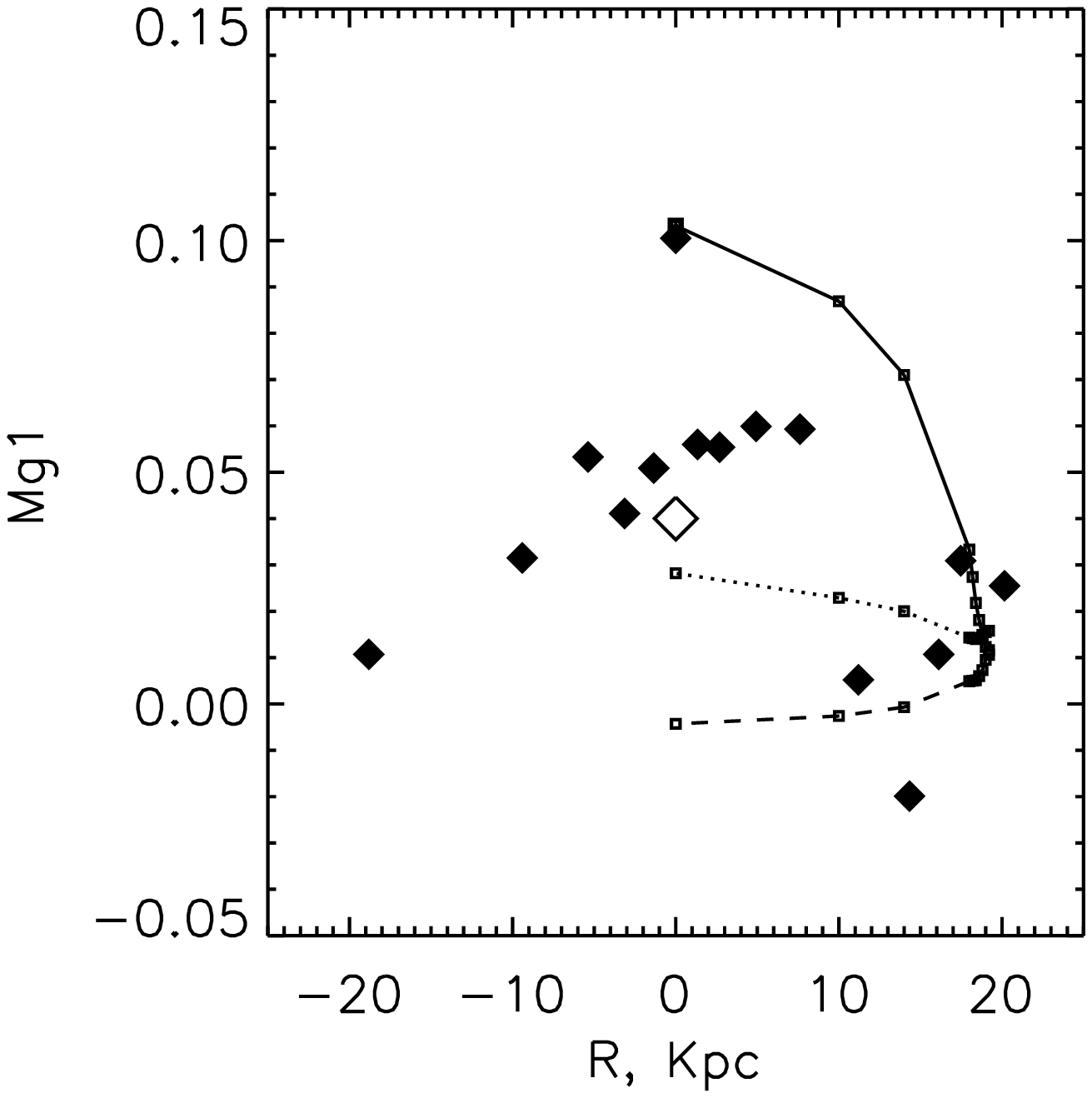}
\includegraphics[width=5cm,angle=0]{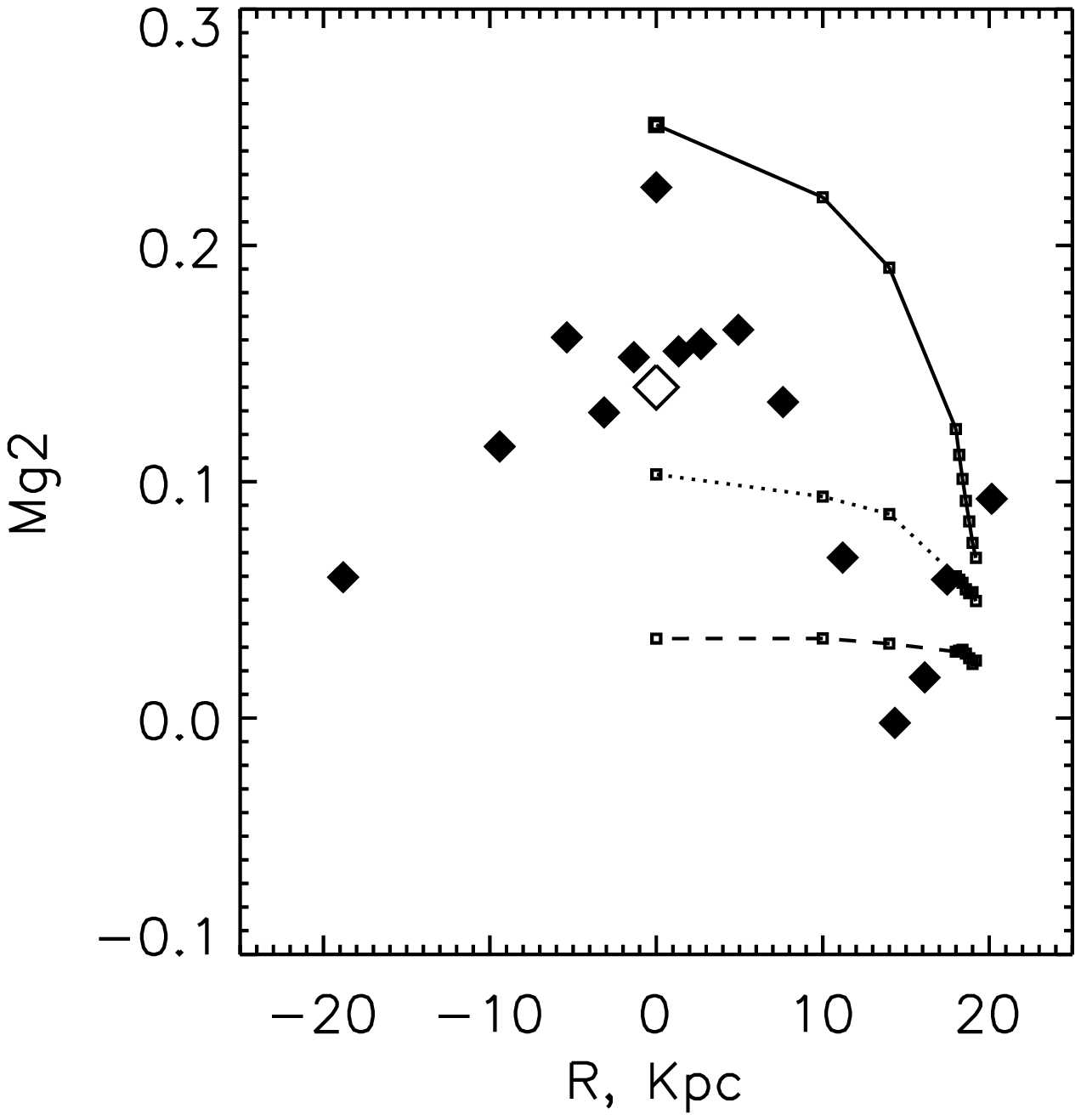}
\includegraphics[width=5cm,angle=0]{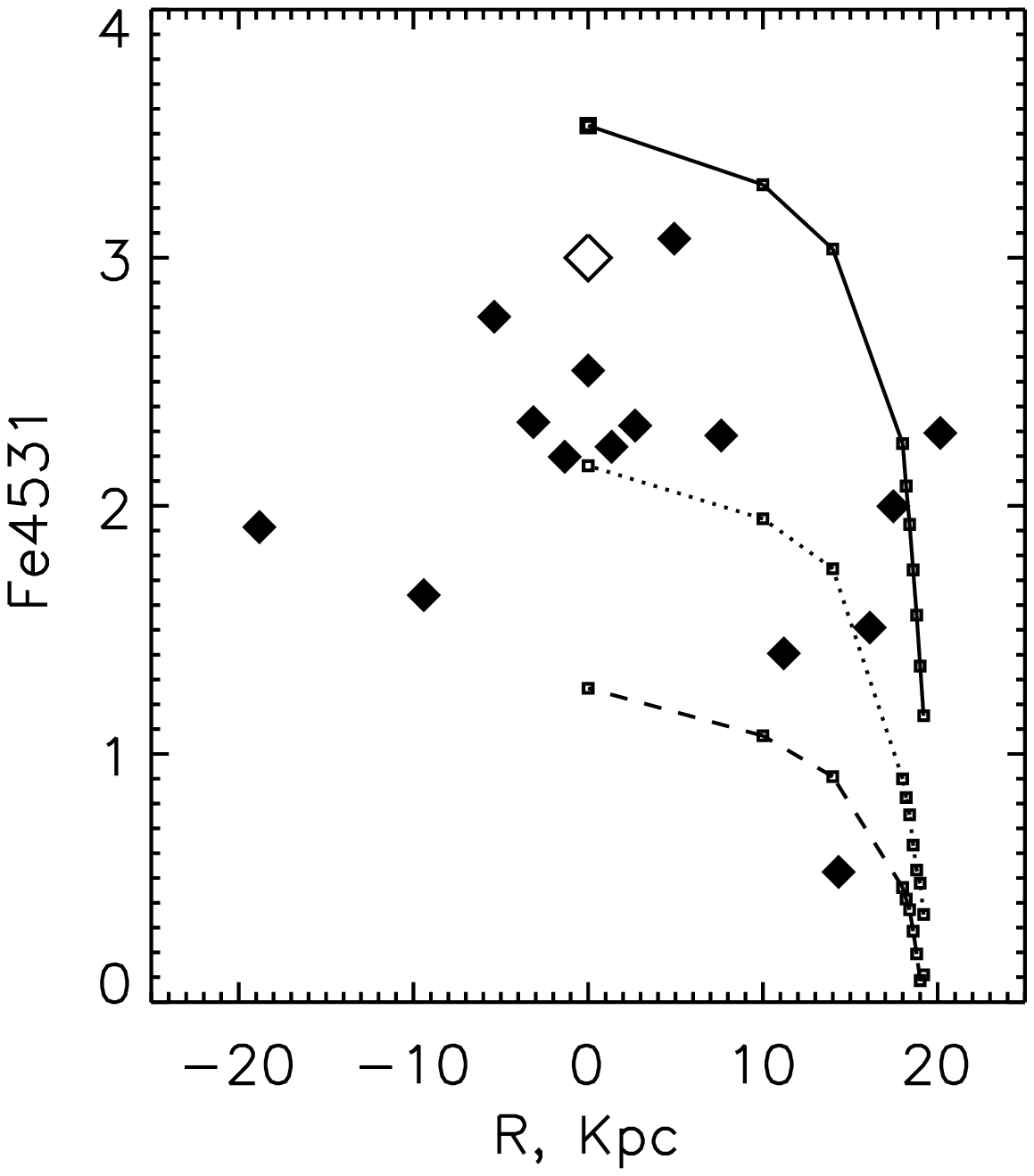}}
 \caption{ Lick indexes  $Mg_1$, $Mg_2$ and $Fe4531$ at different distances from the center of Arp~10
(solid squares). The curves correspond to distribution of the index for the
model single-burst populations of the ages 10, 5, 3, 1, 0.9, 0.8, 0.7, 0.6,
0.5 and 0.4 Gyr and metallicities [Fe/H]=0 (solid), -1 (dotted) and -1.7
(dashed). The open diamond designates the intruder (see below)}
\end{figure}

We use the long-slit spectra to determine available Lick indices ([7]). It
enables us to estimate the age and metallicity of stellar population along the
radius of the galaxy. The filled diamonds in Fig. 5 show the measured indices
$Mg_1$, $Mg_2$, and $Fe4531$ at different galactocentric distances. The
theoretical indices as predicted by the single burst models (based on [8]) are
plotted in Fig.~5 by curves. The solid, dotted, and dashed curves are obtained
for a single-population model with [Fe/H]=0, -1, and -1.7, repectively. The
filled squares in each curve correspond to a 10, 5, 3, 1, 0.9, 0.8, 0.7, 0.6,
0.5 and 0.4 Gyr old stellar population (from left to right). The x-axis
spacing of the model curves is an arbitrary linear function of the model age.
It is seen that the nucleus of Arp~10 consists of mostly old and abundant
stellar population. Given the metallicity in the outer ring is rather high
(1/2 of solar according to [5]), one can argue that the stellar population is
getting systematically younger toward the edge of the galaxy, and that its age
changes smoothly with the radius. It supports the propagating nature of star
formation in Arp~10. Since the old population is non-negligible in the inner
parts of Arp~10, a more complex modeling is required to estimate the age of
young population there.

\begin{figure}
\includegraphics[width=13cm,angle=0]{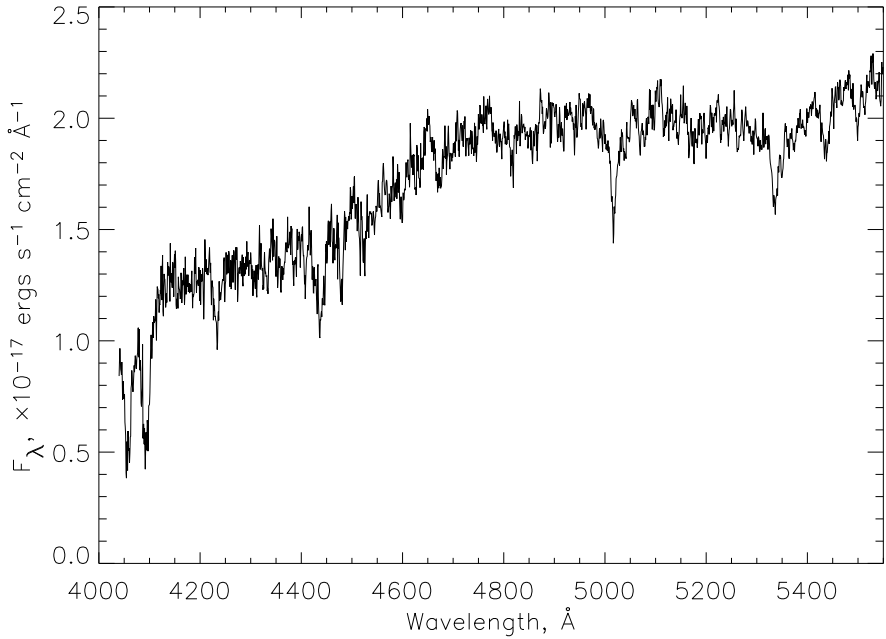}
\caption{Spectrum of the bright knot, the proposed intruder. Hydrogen
absorption lines indicate the presence of significant fraction of A-stars
which reveal a few $10^8$ yrs old population.}
\end{figure}

The spectrum of the bright knot located at $5''$ from the galactic center is
shown in Fig. 6. It provides evidence that the knot is not a bright projected
star, but a dwarf elliptic galaxy located nearby Arp~10. Its radial velocity
is 350 $\km$ higher than that of the main galaxy. The line-of-sight velocity
dispersion is of the order of 100 $\km$. It is comparable with 160 $\km$
dispersion inferred for the center of Arp~10. If the dwarf satellite is
gravitationally bound, its mass should be roughly $10\%$ of that of Arp~10.
Alternatively, and more probably, the satellite has almost been completely
destroyed and absorbed. The Balmer absorption lines (H$\beta$, H$\gamma$,
H$\delta$) are well seen in the spectra what reveals the presence of
significant population of A-stars. Hence, the satellite experienced an episode
of star formation a few hundred million years ago, which corresponds to the
expected time passed since the collision.

\bigskip
{\rm REFERENCES}
\bigskip

\noindent [1] Arp H., 1966, Atlas of Peculiar Galaxies, Pasadena, Caltech

\noindent [2] Appleton P., Marston A., 1997, AJ, 113, 201

\noindent [3] Charmandaris V., Appleton P., Marston A., 1993, Apj 414, 154

\noindent [4] Charmandaris V., Appleton P., 1996, ApJ 460, 686

\noindent [5] Bransford M., Appleton P., Marston A., Charmandaris V., AJ 116, 2757

\noindent [6] Moiseev A.V., 2002, Bull. SAO, 54, 74 (astro-ph/0211104),

\noindent http://www.sao.ru/hq/moisav/scorpio/scorpio.html

\noindent [7] Worthey G., Faber M., Jesus Gonsalez J., Burstein D., 1994, ApJSS 94,
687

\noindent [8] Worthey G., 1994, ApJSS, 95, 107

\end{document}